\documentclass[12pt,a4paper]{article}
\usepackage[utf8]{inputenc}
\usepackage[magyar]{babel}
\makeatletter
\renewcommand\@biblabel[1]{#1.}
\makeatother
\usepackage{graphicx} 
\usepackage[T1]{fontenc}
\usepackage{amsmath}
\usepackage{amsfonts}
\usepackage{amssymb}
\usepackage{graphicx}
\usepackage{floatrow}
\usepackage{palatino}
\usepackage[labelformat=empty]{caption}
\usepackage{t1enc}
\begin{document}

\, \,
\vspace*{5mm}

\begin{center}
{\bf{Symmetries in Stellar, Galactic, and Extragalactic Astronomy}}\\

\bigskip

L\'aszl\'o Szabados$^{1,2}$\\

\bigskip

\end{center}

\noindent $^1$  Konkoly Observatory, Research Centre for Astronomy and Earth Sciences, Konkoly Thege Mikl\'os \'ut 15-17, H-1121 Budapest, Hungary\\

\noindent $^2$ MTA CSFK Lend\"ulet Near-Field Cosmology Research Group\\

\noindent {\it E-mail:} szabados@konkoly.hu\\

\noindent {\it ORCID: 0000-0002-2046-4131}\\


\noindent {\bf Abstract:} Examples are presented for appearance of geometric symmetry in the shape of various astronomical objects and phenomena. Usage of these symmetries in astrophysical and extragalactic research is also discussed.\\


\bigskip

\noindent {\bf{INTRODUCTION}}\\

\noindent Symmetry surrounds us in an ubiquitous manner. This is true for the everyday life and for the vast depths of the Universe, as well. Astronomical observations have revealed various types of symmetry, and these features provide astronomers with useful pieces of information on structure and evolution of various celestial bodies and their groups as well as clusters. This paper offers an insight into opportunities of using symmetry in astrophysical research including Galactic and extragalactic  studies by presenting various examples taken from the current astronomical literature. \\

\bigskip

\noindent {\bf{1 SYMMETRIES IN ASTRONOMY -- HISTORICAL CONTEXT}}\\

\noindent Our Universe is full of symmetries and their violations. In ancient times the planets were thought to move along a circle and the extraterrestrial realm consisted of several embedded spheres -- the sphere and the circle being most perfect 3D and 2D shapes, respectively. The extended objects visible with the naked eye (Sun and Moon) seem to be also circular. Belief in perfection in structure of the Universe and motions in it dominated the view about cosmos for millennia. Spherical and circular symmetries seemed to characterise the whole extraterrestrial world.\\

\noindent Copernican revolution, invention of the telescope, and increasing accuracy of astronomical measurements, all contributed to decline of this belief. With emergence of the concept of the heliocentric system, Earth lost its central position in the Universe, and Kepler's laws of planetary motion, later generalised by Newton's laws of dynamics have shown that motion of a celestial body along circular trajectory is an extremely rare exception. Nevertheless, ellipse is also a symmetric configuration, though its symmetry is not as perfect as for a centrally symmetric circle.\\

\noindent Although symmetry is an ancient notion, it is worthy to note that in the works written by Copernicus and Galileo symmetry was not used according to its modern meaning, instead it referred to some sense of proportionality (Han \& Goldstein, 2004).\\

\noindent Recently, owing to further development of astronomical observational technique (giant telescopes, high sensitivity imaging detectors, use of interferometry allowing extremely high angular resolution, etc.), the shape of the extended celestial bodies and contours of their neighbourhood can be studied in a detailed manner. The opportunity to get an image of various astronomical objects and phenomena, however, directed only a few observer's attention to study symmetries from the point of view of astrophysics. Even more surprising that the most comprehensive summary on astrophysical symmetries, prepared by Trimble (1996) has remained uncited up to now.\\

\noindent Nevertheless, there are occasional studies in the astrophysical literature in which symmetry and its role are key elements. Some of these research areas are mentioned in this paper to emphasize importance of symmetries in understanding various astrophysical phenomena.\\

\noindent The relationship between astronomy and symmetry studies is, however, much more diversified. It includes, e.g. the problem of parity breaking: the amount of matter and antimatter widely differs from each other, for the Universe is matter-dominated. The deficit of anti-matter is, however, a favourable circumstance on local scale because interaction of matter and anti-matter particles is an evil phenomenon for the terrestrial life.\\

\noindent Investigations of particular celestial mechanical problems also involve aspects of symmetry: quite recently, K\H{o}v\'ari \& \'Erdi (2020) studied axisymmetric central configurations of the four-body problem with three equal masses. In the same special issue of the journal Symmetry, large scale symmetric structures of some optically selected quasars have been studied at GHz radio frequencies (Krezinger et al. 2020).\\

\bigskip

\noindent {\bf{2 SYMMETRIC CELESTIAL BODIES AND PHENOMENA}}\\

\noindent {\bf{2.1 Spherical symmetry in modern astronomy}}\\

\noindent The shape of non-rotating celestial objects is spherical, if they are sufficiently massive for gravity to dominate over other forces and enough time has elapsed for becoming insensitive for initial conditions. Therefore all slowly rotating stars are essentially spherical. Moreover, old globular clusters, i. e. those star clusters that contain about a hundred thousand members or more fill a spherical region (Figure 1), similarly to most massive clusters of galaxies which are not in gravitational interaction with neighbouring galaxy clusters (Figure 2).\\
\begin{center}
\begin{figure}[!]
\floatbox[{\capbeside\thisfloatsetup{capbesideposition={right,bottom}}}]{figure}[\FBwidth]
{\includegraphics[width=0.6\textwidth]{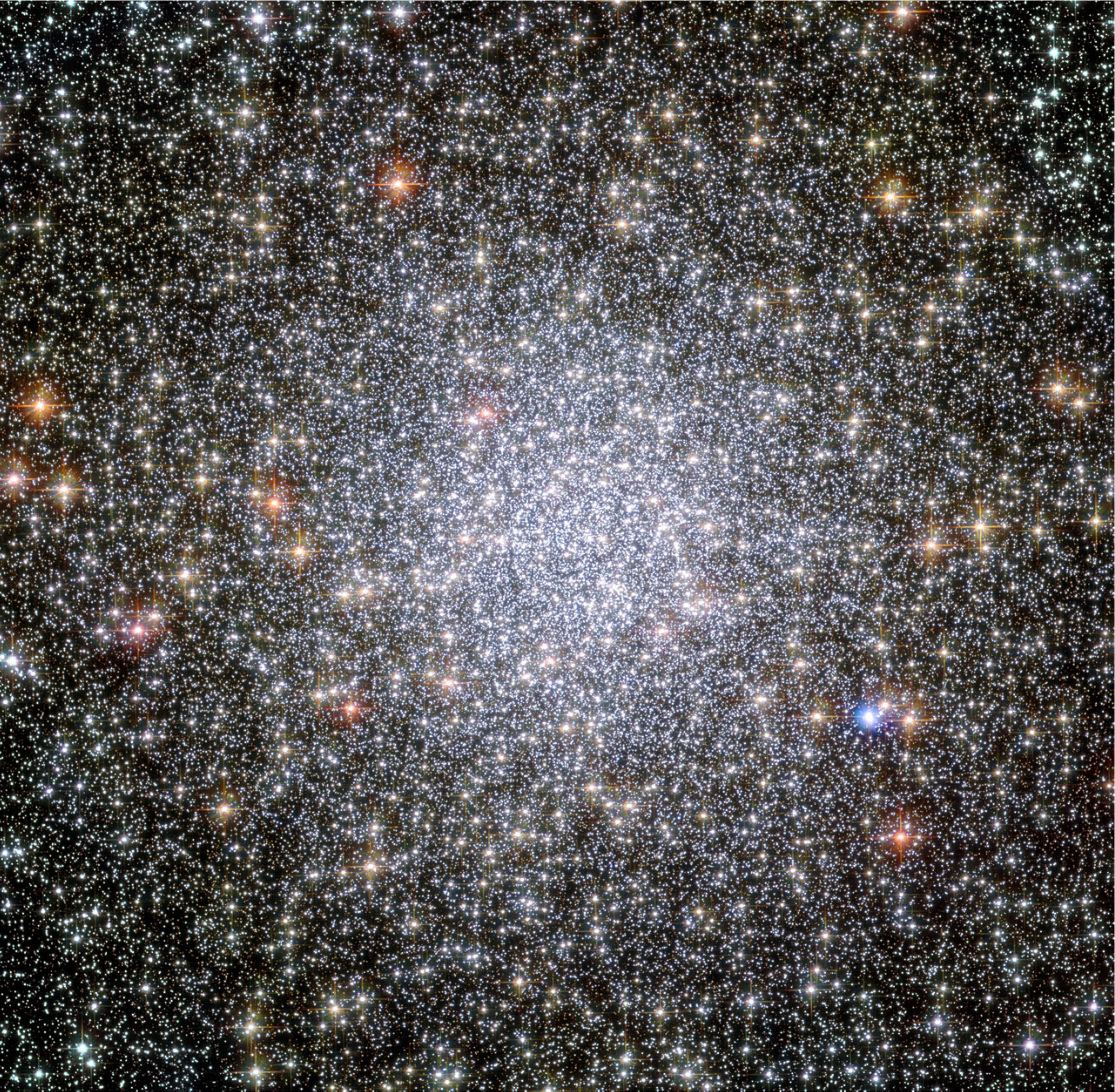}}
{\caption{\footnotesize Figure 1. Globular cluster 47 Tucanae. Credit: NASA, ESA, and the Hubble Heritage Team (STScI/AURA), J. Mack (STScI) \& G. Piotto (University of Padova)}}
\end{figure}
\end{center}

\begin{center}
\begin{figure}[!]
\floatbox[{\capbeside\thisfloatsetup{capbesideposition={right,bottom}}}]{figure}[\FBwidth]
{\includegraphics[width=0.6\textwidth]{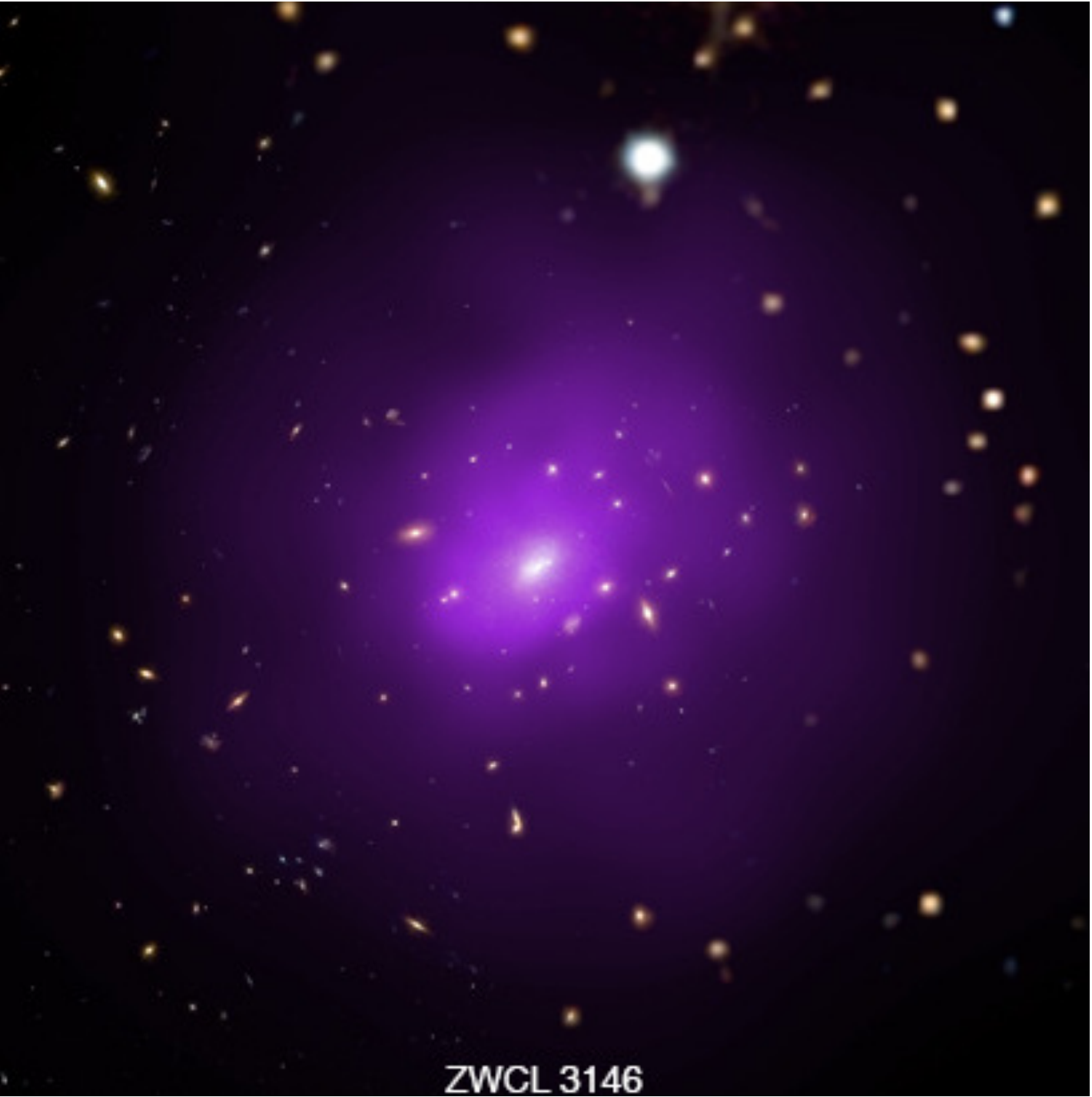}}
{\caption{\footnotesize Figure 2. X-ray image superimposed on the optical image of the galaxy cluster ZWCL 3146. Credit: X-ray: NASA/CXC/Univ. of Alabama/A. Morandi et al.; Optical: SDSS, NASA/STScI.}}
\end{figure}
\end{center}
\vspace*{-20mm}
\noindent The shape of a galaxy cluster can best be studied by x-ray imaging because the hot intergalactic plasma radiating in high energy region of the electromagnetic spectrum pervades the whole cluster.\\

\noindent Our Universe itself is also nearly spherical as seen from our vantage point. This perfect symmetry is testified by isotropy of the cosmic microwave background radiation, as well as by the sky distribution of the observed gamma-ray bursts of cosmological origin.\\

\noindent The celestial world is far from being static. Motions of and interactions between celestial bodies tend to destroy or at least decrease the existing symmetries. Moreover, dynamical phenomena can occur inside isolated stars, as well. Such dynamical phenomenon is, among others, stellar pulsation.
Because stars are located at tremendous distances from our Solar System, periodic changes in the diameter of the stars cannot be detected directly (although a few exceptions based on interferometric observations do exist). A special case leading to spherically symmetric motion inside stars is the radial pulsation of special types of variable stars. Pure radial pulsation can be excited and maintained on sufficiently long time-scale in Cepheid and RR Lyrae type variables (Percy, 2007). This kind of pulsation corresponds to the eigenfrequency of the free oscillation of the given star or some low overtone of the fundamental mode. The pulsation can be studied and characterised by following and analysing periodic variations in brightness and radial velocity of the relevant stars.\\

\noindent Recently, however, it turned out that low-amplitude non-radial oscillation modes can be superimposed on the large amplitude radial pulsation, thus destroying spherical symmetry of the phenomenon.\\
\bigskip

\noindent {\bf{2.2 Celestial objects/phenomena with lesser degree of symmetry}}\\

\noindent There are, however, directly visible manifestations of various degrees of symmetry in astronomical phenomena. In what follows, some typical examples are mentioned.\\

\noindent A shocking symmetrical shape is formed already during the early phases of star formation when a bipolar outflow of matter leaves the newborn star in the form of a pair of collimated jets (Figure 3).\\
\begin{center}
\begin{figure}[!]
\floatbox[{\capbeside\thisfloatsetup{capbesideposition={right,bottom}}}]{figure}[\FBwidth]
{\includegraphics[width=0.6\textwidth]{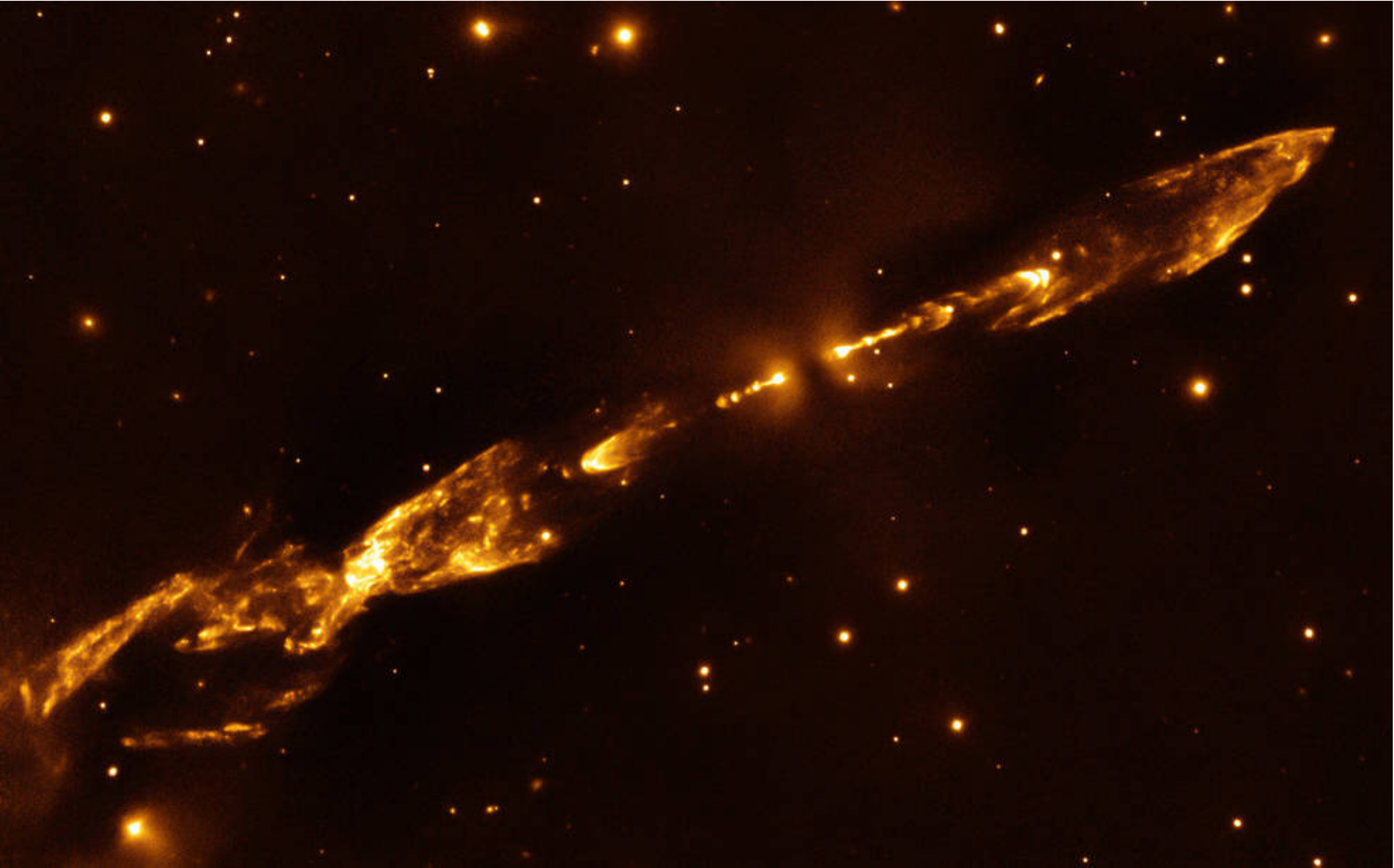}}
{\caption{\footnotesize Figure 3. The Herbig-Haro object HH212 is excited by the bipolar outflow from the central star invisible in this image. Credit: ESO, M. Caughrean.}}
\end{figure}
\end{center}

\vspace*{-10mm}

\noindent Upon impact of the high-speed gas (plasma) with the ambient interstellar medium a shock front is generated leading to visible emission. Though the central star is thought to be spherical, its vicinity shows only axial symmetry.\\

\noindent Planetary nebulae are circumstellar nebulosities formed from the ejected outer layers of low mass stars during late phases of their evolution. These nebulae are excited by intense ultraviolet radiation of the hot central star. Therefore their name is a misnomer, the nebula itself has nothing to do with planets. The name planetary nebula was coined in the late 18th century to these faint round light patches resembling the disk of a planet as seen through the eyepiece of contemporary telescopes available centuries ago. Modern telescopes and imaging techniques, however, revealed the fine details in the structure of these nebulosities.\\ 
\begin{center}
\begin{figure}[!]
\floatbox[{\capbeside\thisfloatsetup{capbesideposition={right,bottom}}}]{figure}[\FBwidth]
{\includegraphics[width=0.6\textwidth]{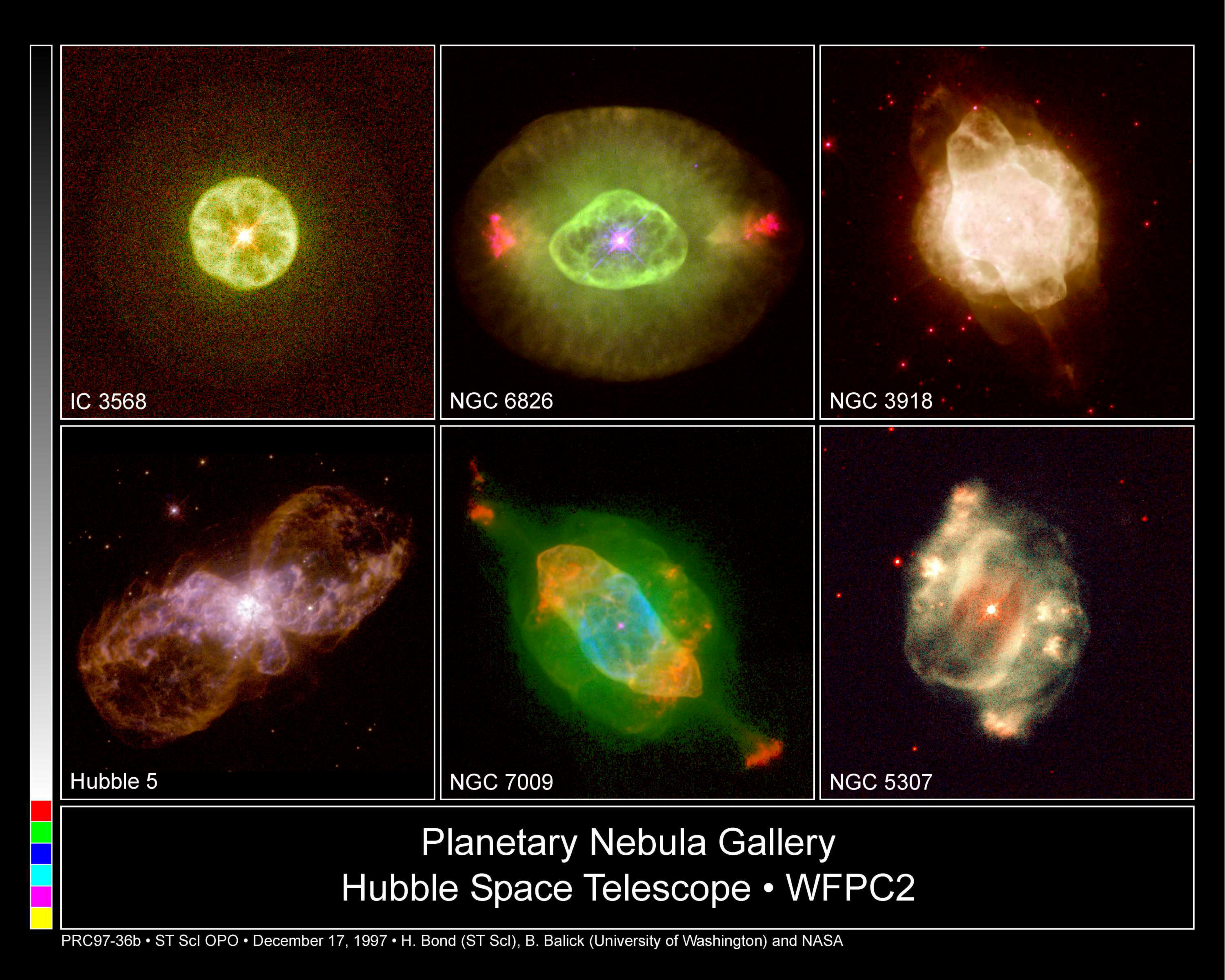}}
{\caption{\footnotesize Figure 4. Gallery of planetary nebulae as seen by the Hubble Space Telescope. The only perfectly symmetrical  one among them is IC 3568 in the upper left corner of the montage. All others show only axial symmetry. Credit: HST/NASA/ESA/STScI, H. Bond \& B. Balick}}
\end{figure}
\end{center}
\vspace*{-10mm}
\noindent Although there exist planetary nebulae of globular shape, most of them lack spherical symmetry, instead their shape is axially symmetric (Figure~4). There are various causes of such aspherical morphology:\\
- {\it The central star is not solitary}, instead, a binary star is located in the centre of the nebula excited by the hotter component and the interaction between the two stars is responsible for generating axial symmetry to the shape of the surrounding nebulosity.\\
- {\it Existence of magnetic field.} Since magnetic monopoles do not exist, an appreciable magnetic field obviously destroys spherical symmetry of the objects it pervades. In electrically charged matter (with infinite electric conductivity) the magnetic field is frozen into the plasma. This means that particles can move along with the local field lines. For the perpendicular motions of the plasma, the field lines will push the matter.\\
- {\it Stellar rotation.} This feature was already mentioned in Section 2.1.
As a result of the underlying physical processes, quite strange shapes of nebulosities can be observed. Figure 5 shows the Red Rectangle, a so called protoplanetary nebula.\\
\begin{center}
\begin{figure}[!]
\floatbox[{\capbeside\thisfloatsetup{capbesideposition={right,bottom}}}]{figure}[\FBwidth]
{\includegraphics[width=0.6\textwidth]{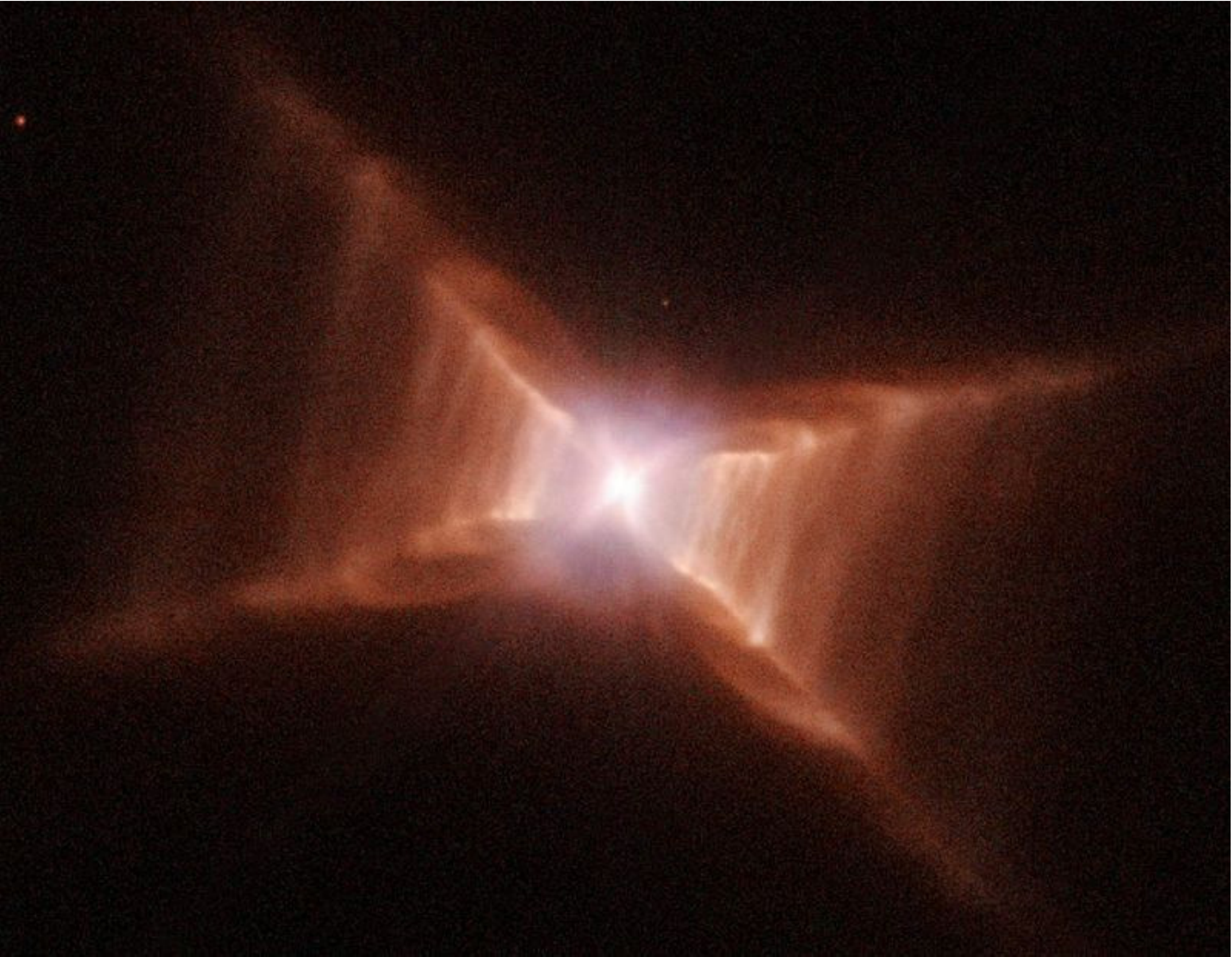}}
{\caption{\footnotesize Figure 5. Red Rectangle, a protoplanetary nebula. The dust torus surrounding the central star (in fact, a binary star) pinches the otherwise spherical outflow into cone shapes which seem to form an X because the torus is seen edge-on. Credit: NASA/ESA/HST/STScI/ H. Van Winckel, M. Cohen \& H. Bond.}}
\end{figure}
\end{center}
\vspace*{-10mm}
\noindent Another kind of spectacular nebulosity occurs in expanding supernova remnants formed as a consequence of a cataclysm during the final stage of the evolution of intermediate mass stars, In ideal case this supernova explosion is centrally symmetric, so the shape of the remnant is expected to be spherical, resulting in a circular shape projected on the sky. Observations, however, testify that many supernova remnants are lopsided, showing only mirror symmetry referring to the strange fact that the explosion of the dying star itself lacks spherical symmetry (Figure 6). Here, again, binarity of the progenitor may play an important role.\\
\begin{center}
\begin{figure}[!]
\floatbox[{\capbeside\thisfloatsetup{capbesideposition={right,bottom}}}]{figure}[\FBwidth]
{\includegraphics[width=0.6\textwidth]{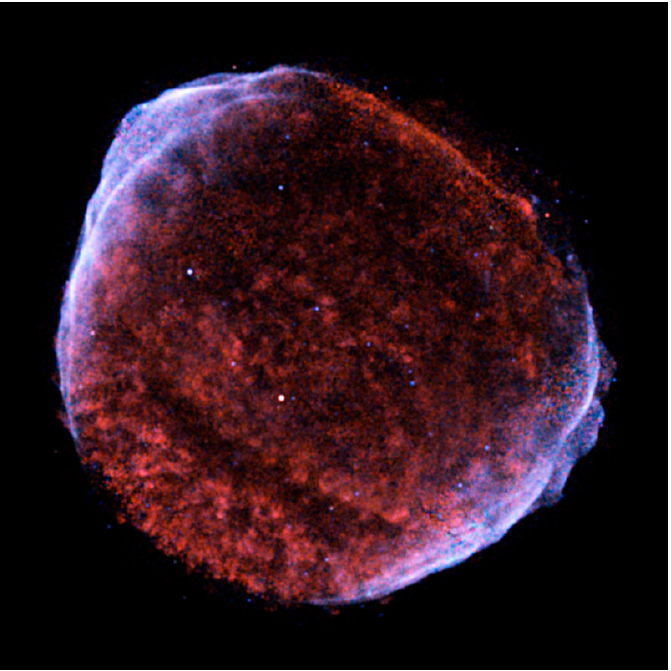}}
{\caption{\footnotesize Figure 6. SN1006, the remnant of the supernova exploded in the year 1006. This x-ray image was obtained with the Chandra X-ray Observatory. Credit: NASA/CXC/P. Frank Winkler}}
\end{figure}
\end{center}
\vspace*{-10mm}
\noindent As to stellar groupings, there are much larger formations than star clusters, named galaxies which contain 10$^8$-10$^{11}$ stars and a considerable amount of gas and dust in interstellar space. The shape of galaxies is diverse: there are elliptical, spheroidal, lenticular, spiral and irregular galaxies. Various forms of symmetry can be observed in most galaxies  excepting irregular systems. One of the key galaxy types in the Universe is the spiral galaxy, as demonstrated in an especially beautiful way in Figure 7. Spiral galaxies (including our Milky Way system) are characterised by rotational symmetry, here the axis of symmetry is perpendicular to the galactic plane and crosses the centre of the galaxy. Mirror symmetry is also observed in galaxies if the system is seen edge-on (Figure 8). Being complex systems, compound symmetry (radial translational, rotation, and reflective) is also a characteristic feature of spiral galaxies.\\
\begin{center}
\begin{figure}[!]
\floatbox[{\capbeside\thisfloatsetup{capbesideposition={right,bottom}}}]{figure}[\FBwidth]
{\includegraphics[width=0.6\textwidth]{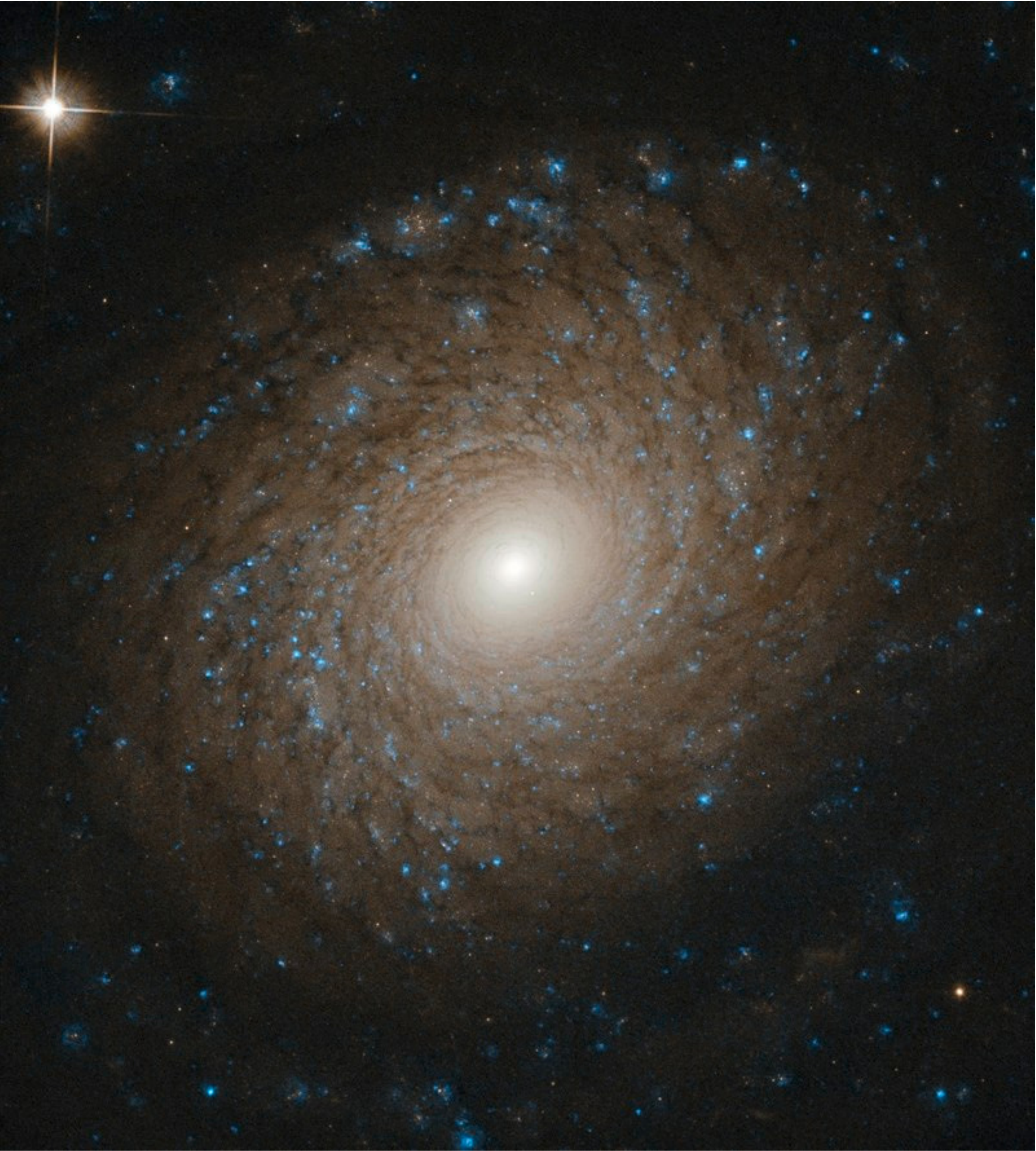}}
{\caption{\footnotesize Figure 7. NGC 2985, an archetypal spiral galaxy with a number of tightly wound spiral arms. Credit: NASA/ESA/HST, L. Ho}}
\end{figure}
\end{center}
\begin{center}
\begin{figure}[!]
\floatbox[{\capbeside\thisfloatsetup{capbesideposition={right,bottom}}}]{figure}[\FBwidth]
{\includegraphics[width=0.6\textwidth]{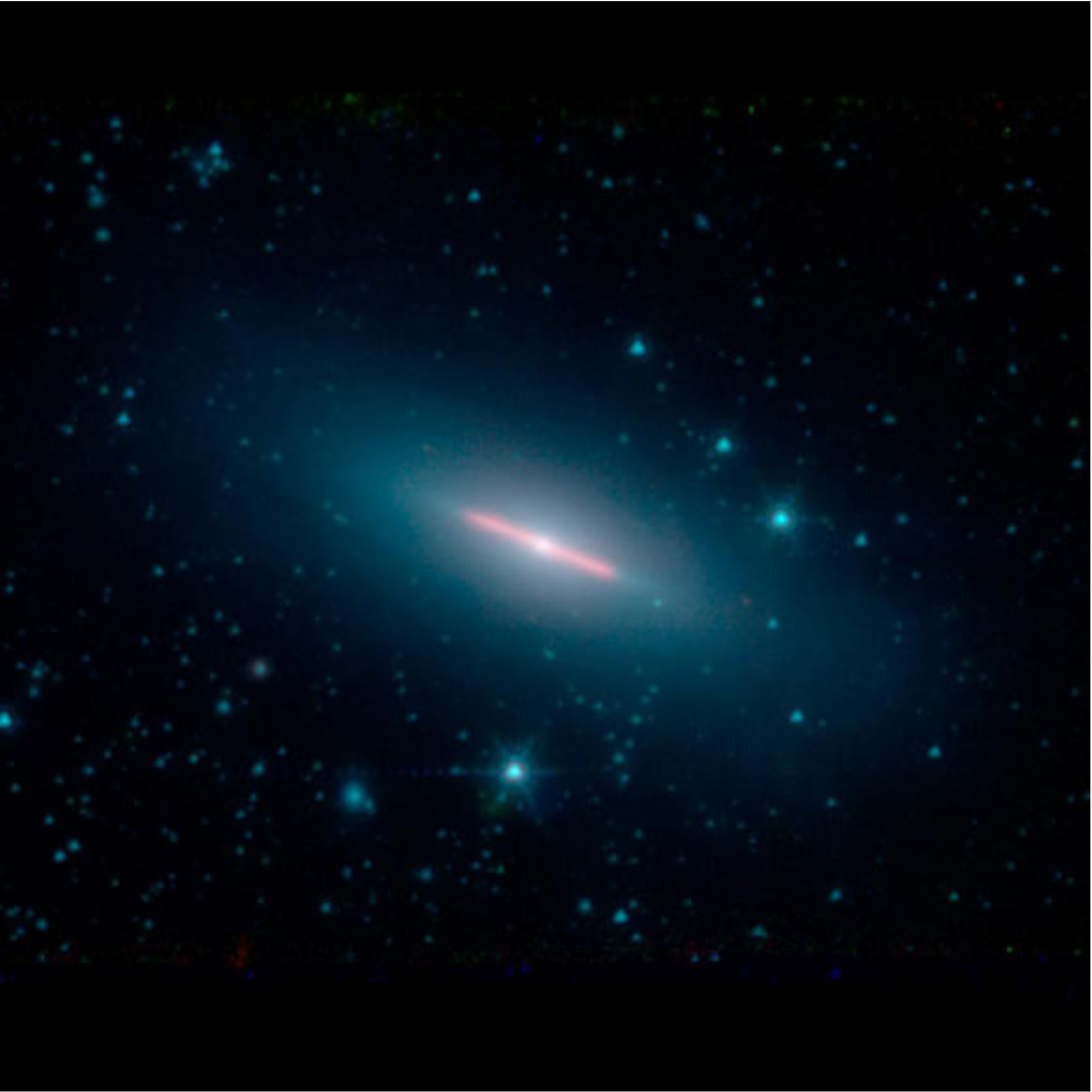}}
{\caption{\footnotesize Figure 8. Near-infrared image of NGC 5866, a distant lenticular galaxy, obtained with the Spitzer Space Telescope. Credit: NASA/JPL-Caltech}}
\end{figure}
\end{center}
\vspace*{-5mm}
\noindent A special case of axial symmetry occurs in the phenomenon of gravitational lensing. If a massive object (star, galaxy) is situated along the line of sight of a more distant object (e.g. galaxy cluster), relativistic light bending distorts the image of the farther object. Depending on the geometry of the involved objects, the image of the farther object will be multiplied (Einstein cross), stretched into a perfect circle (Einstein ring) (Figure 9) or arches of circles (Figure 10).\\
\begin{center}
\begin{figure}[!]
\floatbox[{\capbeside\thisfloatsetup{capbesideposition={right,bottom}}}]{figure}[\FBwidth]
{\includegraphics[width=0.6\textwidth]{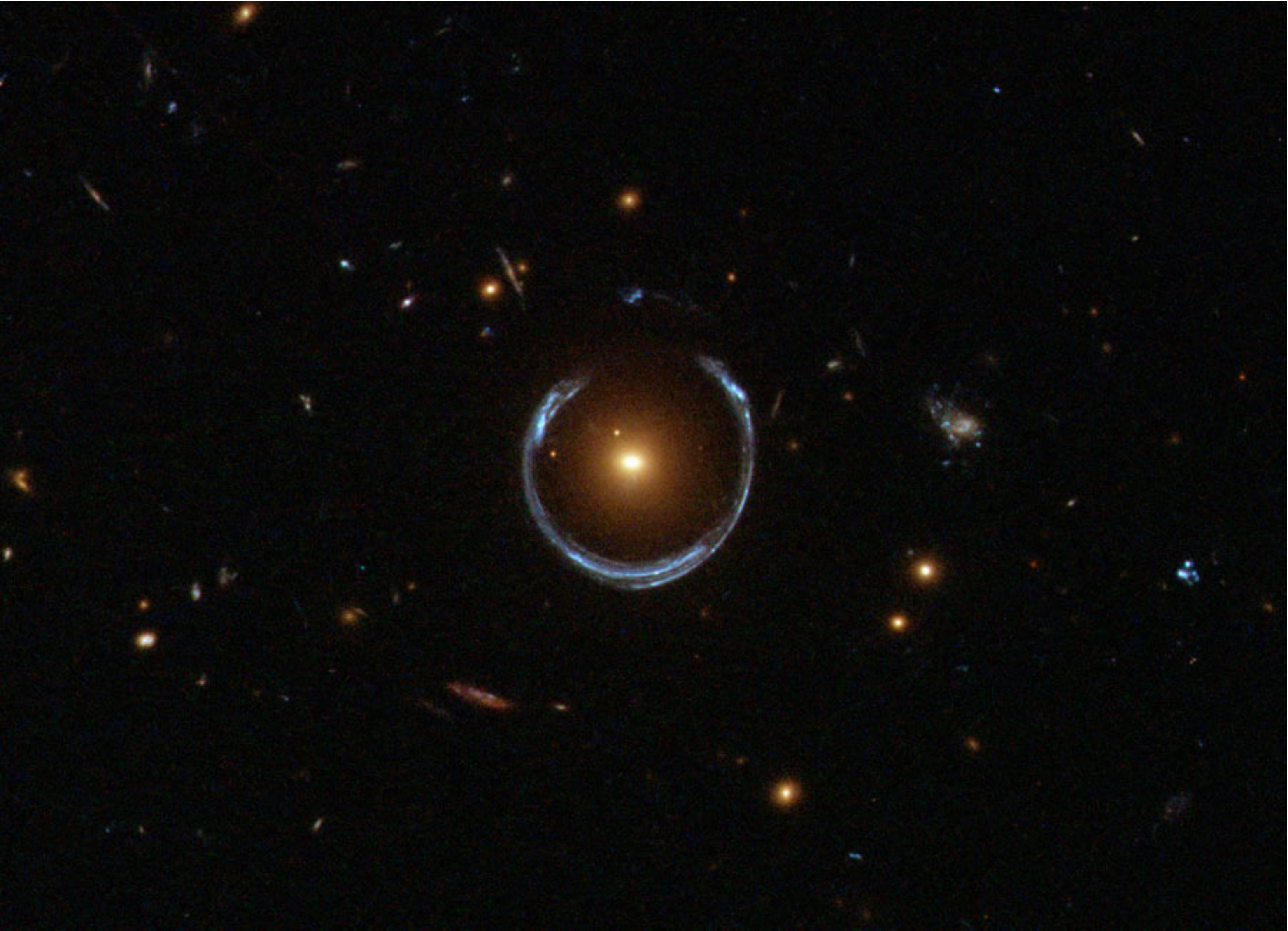}}
{\caption{\footnotesize Figure 9. Gravitational lensing shown by the luminous red galaxy LRG 3-757. Credit: NASA/ESA/HST.}}
\end{figure}
\end{center}
\vspace*{-10mm}
\begin{center}
\begin{figure}[!]
\floatbox[{\capbeside\thisfloatsetup{capbesideposition={right,bottom}}}]{figure}[\FBwidth]
{\includegraphics[width=0.6\textwidth]{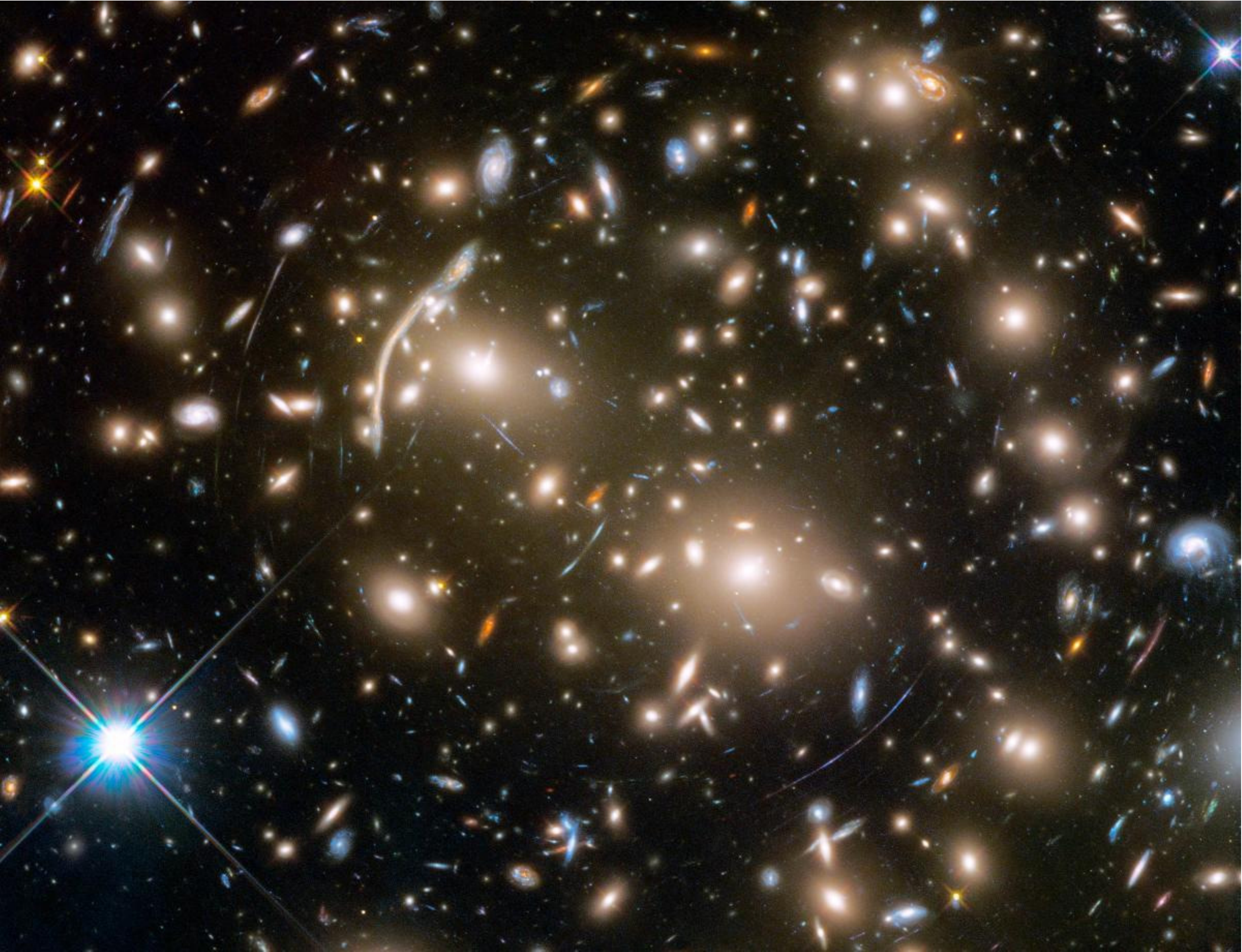}}
{\caption{\footnotesize Figure 10. Gravitational lensing effect caused by the galaxy cluster Abell 370. Credit: NASA, ESA, J. Lotz \& the Hubble Frontier Fields Team (STScI)}}
\end{figure}
\end{center}

\noindent {\bf{3 SCIENTIFIC USE OF THE OBSERVED SYMMETRIES}}\\

\noindent Symmetric natural phenomena and objects -- including astronomical ones -- are aesthetic for human eye and brain. In addition, they provide astronomers with useful pieces of information about physical properties of the involved celestial objects. Below I only mention some examples how the observed symmetries are utilised for astrophysical purposes. In the case of pulsating variable stars, identification of the excited frequencies can help to sound stellar radial structure. This is the topic of a relatively new field of research of asteroseismology.\\

\noindent Recent observational evidence confirms that stars and their planetary systems form simultaneously. This means that each star is surrounded by a disk from which matter is accreted onto the newly formed star and the planets are also formed from the matter in the disk during early phases of stellar evolution. Existence of the disk causes mirror symmetry in the appearance of the protostellar system. Due to the law of angular momentum conservation, the forming star will rotate faster and faster when gathering more matter. To survive this situation, the protostar has to eject some matter to loose its angular momentum. This extra matter cannot leave the star towards equatorial directions because of the circumstellar disk. Therefore the process of angular momentum loss happens perpendicularly to the stellar equator, and this gives rise to high velocity collimated bipolar jets observed in young stars (e.g. Livio, 2004).\\

\noindent Binarity of the involved objects was already mentioned as a possible cause of destroying perfect symmetry. In the solar neighbourhood, frequency of binaries among stars exceeds 50\% (Abt, 1983). Our Sun belongs to minority in this respect. There is growing evidence that the high percentage of planetary nebulae showing only axial symmetry instead of spherical morphology is caused by the fact that the central star has a companion. Even a massive planet orbiting the central star can have a significant effect in destroying spherical symmetry. The importance of binarity in forming and shaping planetary nebulae was recently discussed by Jones \& Boffin (2017). Although rapid rotation of the central star can also cause deviations from spherical symmetry, it cannot lead to the observed axisymmetric bipolar structures because this would require impossibly high rotation rate for a single star. Similarly, magnetic field of the central stars in planetary nebulae is not sufficiently strong to cause the observed shape of nebulosity.\\

\noindent Observations in various bands of electromagnetic radiation (from radio to x-ray) indicate that supernova remnants generally do not possess spherical symmetry which refers to anisotropy of the supernova explosion itself (Meyer et al., 2015). Theoretical modelling of different types of supernova explosions is an important research area in astrophysics of our age. Observed morphology of supernova remnants provides important constraints to validate the models. As an example, properties of the barrell-like (cylindrically symmetric) supernova remnant of SN 1006 shown in Figure 6 were investigated by Bisnovatyi-Kogan et al. (1990).\\

\noindent As far as galaxies are concerned, in addition to axial symmetry shown by spiral galaxies, they are characterised by intrinsic chirality. This feature manifests itself as bimodality of winding of spiral arms with respect to direction of rotation of the whole galaxy. In the case of a leading arm, its tip points in the direction of rotation, while the spiral arm is trailing if its tip points in the opposite direction from the rotation. Chirality of spiral arms has been studied by Capozziello \& Lattanzi (2006) who concluded that winding of arms is in general trailing and the occasionally occurring spiral galaxies with leading spiral arms can be interpreted as a result of tidal effects during galaxy encounters. This feature deserves further investigation.\\

\noindent Even physical properties of the galaxies can be studied based on their observed symmetry. Conselice (1997) investigated a large sample of spiral, lenticular, and elliptical galaxies seen face-on in the following manner: each galaxy image was rotated by 180 degrees and subtracted from the original image to obtain a quantitative value for its structural symmetry. Creating a residual map of all asymmetric components, he thus found a strong correlation between the average colour of the galaxy and the parameter describing symmetry. The explanation of this correlation is that this method is sensitive for ongoing star formation because star forming regions contain several bright blue (high mass) stars. This result shows that symmetry can also be used as a measure of the actual star formation rate.\\

\noindent The use of symmetric features of gravitational lensing is completely different from astrophysical uses discussed above. The radius of the Einstein ring depends on the mass of the lensing object: the more massive it is, the larger the Einstein ring radius. It also depends on the distance between us, the lensing object, and the lensed background source. A careful and detailed observation of individual gravitational lensing phenomena can lead to the determination of the distance of both the lensing object and the lensed source (e.g., Jee et al., 2015). This method of cosmic distance determination can be applied to farthest cosmological objects.\\

\noindent This overview was far from comprehensive, many more astronomical objects and phenomena also show elements of symmetry. Just to mention some of them: mass loss from and nebulosities around Wolf-Rayet stars, Str\"omgren spheres surrounding hot stars, active galactic nuclei, light curves of periodic variable stars, especially eclipsing binaries, appearance and distribution of spots on solar surface during the activity cycle of our star, the Sun.\\

\medskip

\noindent {\bf{ACKNOWLEDGEMENTS}}\\
Preparation of the manuscript was financially supported by the LP2018-7/2020 grant of the Hungarian Academy of Sciences.\\

\medskip

\noindent {\bf{REFERENCES}}\\
Abt, H. A. (1983) Normal and Abnormal Binary Frequencies, Ann. Rev. \hspace*{1cm} {\it Astronomy \& Astrophysics}, 21, 343-372.\\ \hspace*{1cm} https://doi.org/10.1146/annurev.aa.21.090183.002015\\
Bisnovatyi-Kogan, G. S., Lozinskaya, T. A., \& Silich S. A. (1990) Barrell-like \hspace*{1cm} Supernova Remnants, {\it Astrophysics and Space Science}, 166, 277-287.\\ \hspace*{1cm} https://doi.org/10.1007/BF01094899\\
Capozziello, S., Lattanzi, A. \hfill (2006) \hfill Spiral \hfill Galaxies \hfill as \hfill Chiral \hfill Objects,\\ \hspace*{1cm} {\it Astrophysics and Space Science}, 301, 189-193.\\ \hspace*{1cm} https://doi.org/10.1007/s10509-006-1984-6\\
Conselice, C. J. (1997) The \hfill Symmetry, \hfill Color, \hfill and \hfill  Morphology \hfill of\\ \hspace*{1cm} Galaxies, {\it Publications of the Astronomical Society of the Pacific}, 109, \hspace*{1cm} 11, 1251-1255.\\ \hspace*{1cm} https://doi.org/10.1086/134004\\
Hon, G., Goldstein B. R. \hfill (2004) \hfill  Symmetry \hfill  in \hfill  Copernicus \hfill  and \hfill  Galileo,\\ \hspace*{1cm} {\it Journal for the History of Astronomy}, 35, 3, 273-292.\\ \hspace*{1cm} https://doi.org/10.1177/002182860403500302\\
Jee, I., Komatsu, E., \& Suyu, S. H. (2015) Measuring Angular Diameter \hspace*{1cm} Distances of Strong Gravitational Lenses, {\it Journal of Cosmology and \hspace*{1cm} Astroparticle Physics}, 11, ID33.\\ \hspace*{1cm} https://doi.org/10.1088/1475-7516/2015/11/033\\
Jones D., Boffin, H. M. J. (2017) Binary Stars as the Key to Understanding \hspace*{1cm} Planetary Nebulae, {\it Nature Astronomy}, 1, 0117.\\ \hspace*{1cm} https://doi.org/10.1038/s41550-017-0117\\
K\H{o}v\'ari, E., \'Erdi, B. (2020) The Axisymmetric Central Configurations of the \hspace*{1cm} Four-Body Problem with Three Equal Masses, {\it Symmetry}, 12, 648, 1-\hspace*{1cm} 12.\\ \hspace*{1cm} https://doi.org/10.3390/sym12040648\\
Krezinger, M., Frey, S., Paragi, Z., \& Deane, R. (2020) High-Resolution \hspace*{1cm} Radio \hfill Observations \hfill of \hfill Five \hfill Optically \hfill Selected \hfill Type 2 \hfill Quasars,\\ \hspace*{1cm} {\it Symmetry}, 12, 527, 1-13.\\ \hspace*{1cm} https://doi.org/10.3390/sym12040527\\
Livio, M. (2004) Astrophysical Jets, Baltic Astronomy, 13, 273-279.\\ \hspace*{1cm} https://doi.org/10.1073/\\
Meyer, D. M.-A., Langer, N., Mackey, J., Velazquez, P. F., \& Gusdorf, A. \hspace*{1cm} (2015) \hfill Asymmetric \hfill supernova \hfill remnants \hfill generated \hfill by \hfill Galactic,\\ \hspace*{1cm} massive runaway stars, {\it Monthly Notices of the Royal Astronomical \hspace*{1cm} Society}, 450, 3080-3100.\\ \hspace*{1cm} https://doi.org/10.1093/mnras/stv898\\
Percy, J. R. (2007) {\it Understanding Variable Stars}, Cambridge: Cambridge \hspace*{1cm} University Press, xxi, 350 p.\\ \hspace*{1cm} https://doi.org/10.1017/CBO9780511536489\\
Trimble, V. (1996) Astrophysical Symmetries, {\it Proc. Natl. Acad. Sci.}, 93, \hspace*{1cm} 14221-14224.\\ \hspace*{1cm} https://doi.org/10.1073/pnas.93.25.14221

\end{document}